\documentclass[twocolumn,showpacs,preprintnumbers,amsmath,amssymb]{revtex4}
\usepackage{graphicx}
\usepackage{dcolumn}
\usepackage{bm}

\begin{document}

\title{Many body quantum physics in XANES of highly correlated materials, mixed valence oxides and high temperature superconductors}

\author{Antonio Bianconi$^{1,2,3}$}

\affiliation{$^1$ RICMASS, Rome International Center for Materials Science Superstripes, Via dei Sabelli 119A, 00185 Rome, Italy}
\affiliation{$^2$ Solid State and Nanosystems Physics,  National Research Nuclear University, MEPhI, Moscow Engineering Physics Institute,
       Kashirskoye sh. 31, Moscow 115409, Russia}
\affiliation{$^3$ CNR, Institute of Crystallography, Consiglio Nazionale delle Ricerche, via Salaria, 00015 Monterotondo, Italy}


\begin{abstract}

The x-ray absorption near edge structure (XANES), developed in these last 40 years using synchrotron radiation, is a unique tool probing electronic correlations in complex systems via quantum many body final state effects.  Multi electron excitations have been observed first in the sixties in x-ray absorption spectra of atoms and later in molecules and solids. The applications of XANES many body final states to probe unique features of electronic correlation in heavy fermions, mixed valence systems, mixed valence oxides and high temperature superconductors are discussed.

\end{abstract}


\maketitle

\section{Introduction.}

The origin of quantum many body physics is usually associated with the 1926 publication of Heisenberg 
paper \cite{Heise} where he provided the theoretical interpretation of the energy splitting of the spectral
 lines, called ortho- and para- helium, of helium atomic gas absorption spectra.
In this work Heisenberg assumed that the motion of the two electrons in helium is correlated considering it as
the simplest many body electronic system made of only two electrons.
Therefore he described a quantum resonance between two electronic configurations of the two electrons.
In this work he introduced the exchange interaction giving the energy splitting between symmetric 
and antisymmetric wave-functions.
Following the Fermi (1926) article  \cite{fermi} on the Fermi statistics for the electron gas in metals, Dirac (1926) \cite{Dirac}
introduced the distinction of the Fermi versus Bose statistics for fermions  versus bosons characterized by
antisymmetric versus symmetric wave-function respectively. The exchange interaction between 
bosons is attractive, giving at T=0 a quantum Bose condensate,
while the exchange interaction is repulsive for fermions giving at T=0 the degenerate Fermi gas.
Wentzel (1927) \cite{Wentzel} proposed a non trivial extension of the Heisenberg (1926) theory \cite{Heise} to explain the Auger effect.

\section{many body effects in atomic physics}

Majorana (1931) \cite{majorana} extended the Wentzel theory focusing on the selection rules for the non-radiative decay of two electron excitations observed in atomic absorption spectra (Foote \textit{et al.} 1925) \cite{Foote}, (Shenstone, 1931) \cite{Shenstone}.
Following the experiment of Beutler (1935)  \cite{Beutler} on vacuum ultraviolet absorption spectra of rare gases where  broad lines 
due to multi-electron excitations were observed beyond the ionization potential, Ugo Fano (Fano, 1935) \cite{Fano} developed the theory of configuration interaction between discrete and continuum channels with the prediction of the asymmetric line-shape of absorption spectrum of two electron excitations. 

The Fano theory was tested by  the experimental line-shape of shape resonances in nuclear scattering experiments due to configuration interaction between open and closed scattering channels (Blatt \& Weisskopf, 1952)   \cite{Blatt1}, (Blatt \textit{et al.},1953) \cite{Blatt2}, (Feshbach \textit{et al.}, 1954) \cite{Fes1} . This theory was extended to many different scattering channels in many body systems by Feshbach (1958) \cite{Fes2}. The Fano-Feshbach resonances are today of high scientific interest in many different fields and in particular for ultracold gases and high-$T_c$ superconductivity (Vittorini-Orgeas \&  Bianconi, 2009)  \cite{Vittorini}.

Ugo Fano at National Bureau of Standard in 1959 promoted synchrotron radiation research using a 180 MeV synchrotron to measure high resolution soft x-ray absorption spectroscopy of atoms. He proposed to investigate the line-shape of two electron excitations, and published a second extended version  of his 1935 paper in Physical Review (Fano 1961) \cite{Fano2}. 
The experiments  (Madden \&  Codling, 1963) \cite{Madden1}, (Madden \&  Codling, 1965) \cite{Madden2}  provided a clear experimental confirmation of the asymmetric Fano line-shape   
for two electron excitations in helium, neon and argon where the interference of a discrete autoionized state with a continuum gives rise to characteristically asymmetric peaks in soft x-ray absorption spectra (Cooper \textit{et al.}, 1963) \cite{Cooper}. 

The Fano-Feshbach resonance in many body configuration interaction between discrete and continuum states is a fundamental quantum process first observed in x-ray absorption spectra but it was soon applied in condensed matter physics to describe magnetic impurities embedded in a metallic host, in the so-called Anderson Impurity Model  (Anderson, 1961) \cite{Anderson} and in the prediction of the characteristic change in electrical resistivity with temperature known as Kondo effect  (Kondo, 1964) 
\cite{Kondo} which is due to the scattering of conduction electrons by magnetic impurities. Today it is very popular in a variety systems ranging from heavy fermions to Kondo insulators. 

The investigations of two-electron excitations have been extended to different excitations  at high energies in helium Dhez \&  Ederer 1973 \cite{dhez}
and in kripton  (Bernieri  \& Burattini 1987) \cite{bernieri}. The two-electron excitations have been found later in solids, as for example in solid neon (Soldatov \textit{et al.}, 1993) \cite{soldatov} in V, Cr, and Mn $3d^0$ compounds with tetrahedral coordination
 (Bianconi \& Garcia \& Benfatto \textit{et al.}, 1991)  \cite{bianconi1991a}
   in  $SiCl_3$ and $SiF_3$ (Di Cicco \textit{et al.},1992) \cite{dicicco}, in hydrated calcium (D Angelo \textit{et al.}, 2004) \cite{dangelo} and in rare earth alloys  (Chaboy,1990) \cite{chaboy1990}, (Chaboy,1994) \cite{chaboy1994}.

\section{Many body effects in condensed matter}

In 1970 the interest of the synchrotron radiation research in novel facilities using small synchrotrons in Madison (USA) and Tokyo (Japan) and large synchrotrons in Hamburg (Germany)  and and Frascati (Italy) shifted from atomic physics to condensed matter.  New many body theories predicted many body final state effects at the absorption threshold of metals in the soft x-ray absorption spectra  (Roulet \textit{et al.}, 1969) \cite{roulet} (Nozieres \& De Dominicis, 1969) \cite{nozieres} (Schotte \&   Schotte, 1969),  \cite{shotte} (Mahan, 1971) \cite{mahan}  (Hedin \textit{et al.}, 1971) \cite{heidin} (Combescot \&  Nozieres, 1971) \cite{combescot}. 
Similar many body effects predicted in x-ray photoelectron spectroscopy (XPS) (Doniach \&  Sunjic, 1970)  \cite{doniach}. for metals have been supported  by the measurements of asymmetric line-shape of core level spectra (Antonangeli \textit{et al.}, 1977) 
 \cite{antonangeli}. . 
Many body final states effects related with the core hole excitation have been observed in the x-ray absorption spectra of palladium hydrides  (Benfatto \textit{et al.}, 1983)   \cite{benfatto1983} and  in photoemission cross-section  near the carbon K shell  threshold, many body plasmon excitations have been shown to play a relevant role (Bachrach \& Bianconi, 1982) 
\cite{bachrach1982}.
However the many body effects in x-ray absorption spectra of metals give only subtle changes of the line-shape of absorption threshold which are difficult to be observed.

\section{Shape resonances in molecules and condensed matter}

The interest of the scientific community has been focussed 
mainly to understand one-electron final states effects due 
to the role of density of states (DOS)  in the conduction band as
 shown for example in the edge region of the $L_2,$$_3$ absorption spectrum of aluminum (Balzarotti \& Bianconi  \& Burattini 1974)  \cite{balz1974} measured at the facility Solidi Roma at the 1 GeV synchrotron in Frascati, Italy.
Synchrotron radiation spectroscopy using high energy and large accelerators (3.5 GeV SPEAR in Stanford, 7.4 GeV Desy in Hamburg) focused on the weak features in the high energy range called x-ray absorption fine structure (EXAFS). The theory and experiments on weak EXAFS oscillation in a variety of complex systems evolved rapidly from 1970 to 1980. On the contrary the theoretical interpretation of the low energy part of the x-ray absorption spectra giving strong absorption absorption peaks in the continuum of solids, both crystals and glasses (Balzarotti \& Bianconi \& Burattini \textit{et al.}, 1974)  \cite{balza1974} 
remained obscure. These peaks were qualitatively assigned to transitions to unoccupied DOS of the crystal or atomic final states. The physics was clarified by the observation of the shape resonance in the continuum of K-edge absorption of the nitrogen molecule
 (Bianconi \& Petersen \& Brown \textit{et al.}, 1978) \cite{BianconiPetersen1978}. 
 
 The shape resonance is a short living final state made of a quasi stationary state, formed by a multiple scattering resonance of the photoelectron, degenerate with the continuum. Its spatial localization in a nanoscale cluster is determined by the high scattering amplitude of the low kinetic energy, 10-50 eV, of the photoelectron from neighbor atoms. The very short life time, of the order of $10^{-15} s$, is determined by the core hole life time and the shape resonance decay into the continuum.  The shape resonances are related with the effects of electronic correlation in molecules 
(Langhoff \& Davidson 1974) \cite{Langhoff1974}, 
(Broad \& Reinhardt 1976) \cite{Broad1976}, 
(Dehmer \& Dill, 1975) \cite{Dehmer1975}, 
(Dehmer \& Dill, 1976) \cite{Dehmer1976}, 
(Dehmer \textit{et al.}, 1979) \cite{Dehmer1979},
(Loomba \textit{et al.}, 1981) \cite{Loomba1981}.  In a set of papers (Bianconi \& Doniach \& Lublin 1978) \cite{BianconiDoniach1978} 
 (Bianconi, 1979) \cite{Bianconi1979}
  (Bianconi, 1980)  \cite{Bianconi1980}
 the acronym  XANES (x-ray absorption near edge structure) was coined to indicate the spectral features in the absorption spectra of condensed matter, including biological systems, chemical compounds, liquids, amorphous systems and glasses due to shape resonances for the photoelectron which in the final state is confined for a short time in a nanoscale cluster centered on the selected absorption atom. The XANES energy range, extending over 50-100 eV beyond the absorption edge, is defined where the photoelectron wavelength is larger that the interatomic distance. Here the full multiple scattering theory is needed to predict the observed multiple-scattering resonances and it was shown that the XANES variation due subtle structural changes in Fe II and Fe III hexacyanide complexes can be predicted and understood
 (Bianconi \& Dell'Ariccia \& Durham \textit{et al.}, 1982) \cite{Bianconi1982}. 

The XANES data analysis (Della Longa \textit{et al.}, 1995)   \cite{DellaLonga1995}
using the full multiple scattering theory provides today the 
coordination geometry and higher order structural correlation function in the liquid phase 
(Garcia \&  Bianconi \&  Benfatto, \textit{et al.}, 1986)  \cite{garciaBianconi1986}, 
(Garcia \&  Benfatto \&  Natoli \textit{et al.}, 1986) \cite{garcia1986}
 and in a large variety of disordered systems.

\section{Mixed valence systems and heavy fermions}

 The search for many body final states in XANES focused on  valence fluctuating materials like Yb inter-metallics  
(Rao  \textit{et al.}, \cite{Rao1980}  1980) and
TmSe (Launois \textit{et al.},1980)  \cite{Launois1980} 
(Bianconi \& Modesti \& Campagna \textit{et al.},1981)    \cite{BianconiModesti1981} 
and $CePd_3$ (Sarode \textit{et al.}, 1981)  \cite{Sarode1981} 
  were it was shown that the quantitative  mixed valence state could be extracted from the data.
  
After the discovery of superconductivity in the first  heavy fermion system $CeCu_2Si_2$ (Steglich \textit{et al.}, 1979) \cite{Steglich1979}  few groups started to look for experimental evidence of strong electronic correlation in intermetallics. Heavy fermions are at the edge between superconductivity and magnetism. In heavy fermions at the chemical potential a narrow f band, with a large effective mass, coexists with a wide d band 
(Jarlborg \textit{et al.}, 1983). \cite{Jarlborg1983} \ These materials show strong Pauli paramagnetism and electronic complexity (Lieke \textit{et al.},1982) \cite{Lieke1982} 
and their electronic properties emerge from multi-electron electronic configurations.
The breakdown of single  electron approximation in the XANES of heavy fermions like $CeCu_2Si_2$ 
(Bianconi \& Campagna \& Stizza \textit{et al.}, 1981) \cite{BianconiCampagna1981}  and 
$EuPd_2Si_2$ \cite{Padalia1981}  has been found and evidence for the competition between two many body electronic configurations was clearly detected by XANES. 

   The intermediate valence probed by core-level spectroscopy  has  been investigated  in light  Ce vs  heavy Tm rare earths. Unique information on electronic correlation, localization and local-environment effects have been obtained by L-edge XANES
 (Bianconi \& Campagna \& Stizza  \textit{et al.}, 1982).   \cite{BianconiCampagna1982}  
The changes of many body electronic configurations at the Fermi level at the phase transition from $\alpha$ to $\gamma$ phase of cerium have been observed at the Ce $L_3$ absorption edge 
(Lengeler \textit{et al.}, 1983) \cite{Lengeler1983}. 
The effects of dilution and chemical pressure on mixed valence cerium-based systems like $CeNi_5$ have been studied by (Raaen \& Parks, 1983) \cite{Raaen1983} (Parks  \textit{et al.}, 1983)  \cite{Parks1983}.
The role of ocalization and mixing of many body configurations in intermetallic compounds $RPd_3$ (R= La, Ce, Pr, Nd, Sm) have been measured by
 (Bianconi \& Marcelli  \& Davoli \textit{et al.}, 1984)    \cite{BianconiMarcelli1984} 
 and (Marcelli \textit{et al.},1984)  \cite{Marcelli1984} 
 (Marcelli, \textit{et al.}, 1985)   \cite{Marcelli1985},
(Bianconi \& Marcelli  \textit{et al.}, 1985) \cite{BianconiMarcelli1985} 
  by $L_3$ and $L_2$-XANES spectra. The many body final states 
  in inner shell photoemission and photoabsorption spectra of La and Ce 
  strongly correlated electronic compounds have been interpreted theoretically using  the Anderson impurity model 
  (Kotani \& Jo \& Okada  \textit{et al.}, 1987) \cite{KotaniJo1987},   
  (Kotani \& Okada \& Jo \textit{et al.}, 1987). \cite{KotaniOkada1987}

\section{Strongly correlated systems: charge transfer Mott insulators and insulator to metal transitions}

While in the x-ray absorption near edge spectroscopy a very high energy photon (in the KeV energy range) is absorbed by a deep core level, the excited electron in the final state occupies the lowest unoccupied electronic states at the chemical potential therefore many body  
electronic configurations ib correlated electronic systems determine the XANES final states. This feature was used to probe the variation of many body electronic configurations at  the metal-insulator transition in $V_2O_3$ (Bianconi \& Natoli,1978)  \cite{BianconiNatoli1978}    
and in $VO_2$ (Bianconi,1982)  \cite{Bianconi1982}  measuring the high resolution vanadium K-photoabsorption spectrum. 
 In 1985 the theory of correlation gaps in the electronic structure of transition-metal compounds 
 was presented (Zaanen \textit{et al.},1985)  \cite{Zaanen1985} showing the existence of different class of Mott insulators involving both the metal and ligand orbitals. 
 Davoli \textit{et al.} in 1986  \cite{Davoli1986}   provided direct evidence for the correlation gap between $3d^8$  and $3d^9L$ multi-electron configurations (where L indicated the hole in the ligand orbital) in the x-ray-absorption near-edge structure of NiO 
  using oxygen K-edge XANES in agreement with predictions of (Zaanen \textit{et al.}, 1985) \cite{Zaanen1985}.
  Many body configurations in the x-ray absorption spectra of charge transfer gap insulators have been measured in other correlated systems like nickel dihalides   (van der Laan  \textit{et al.}, 1986) \cite{van der Laan1986} in agreement with the teory  (Zaanen \textit{et al.}, 1986)  \cite{Zaanen1986},  uranium 5f oxides and glasses, (Petiau, \textit{et al.}, 1986) \cite{Petiau1986},  
  and 4f compounds (Bianconi \& Marcelli \& Dexpert \textit{et al.}, 1987) \cite{BianconiMarcelliDexpert1987}  like $PrO_2$  (Bianconi \& Kotani \textit{et al.}, 1988) \cite{Bianconi Kotani1988}  and in in $Ce(SO_4)_2$ (Bianconi \& Marcelli  \& Tomellini \textit{et al.}, 1985) \cite{BianconiMarcelli1985}. A full information on ground state many body configurations  of $CeO_2$, (M$O_2$ (M=Ce,Pr,Tb,Hf) and of $LaF_3$ has been  obtained by joint Ce 3p XANES and  deep metal 3p core x-ray photoemission spectroscopy XPS using hard x-rays spectroscopy of tetravalent oxides (Bianconi \& Clozza \textit{et al.}, 1989)  \cite{BianconiClozza1989},  (Bianconi \& Miyahara \& Kotani \textit{et al.}, 1989) \cite{BianconiMiyahara1989}.

\section{high temperature superconductors}

The discovery in 1986 of high temperature superconductivity in copper oxides triggered a large interest on the nature of new metallic states induced by chemical doping of strongly correlated layered copper oxides.
The strongly correlated many body state of the copper oxide plane was studied by Cu $L_{2,3}$-XANES (Bianconi \& Congiu-Castellano  \& De-Santis \& Rudolf, \textit{et al.}, 1987)  \cite{BianconiRudolf1987}, by Cu $K$-edge XANES (Bianconi \& Castellano \& De Santis  \& Politis \textit{et al.}, 1987)  \cite{BianconiPolitis1987}  and by Cu 2p x-ray photoelectron spectroscopy (Bianconi \& Congiu-Castellano \&  De-Santis \&  Delogu \textit{et al.}, 1987) \cite{BianconiDelogu1987}.
These experiments provided the first direct measure of the Mott charge transfer gap between the many body configurations Cu$3d^9$-O$2p^{10}$  (called $3d^9$) and Cu$3d^{10}$-O$2p^5$ (called $3d^{10}$L where L is the oxygen hole) in the parent compounds which was measured to be about 2 eV. These systems can be turned from insulator to metal by variable oxygen interstitials content y like in the high $T_c$ superconductor $YBa_2Cu_3O_{6+y}$. The variation of the many body final states in XANES with variable oxygen content  provided the first direct experimental evidence that the itinerant Cu$3d^9$L (where L is the hole on oxygen orbital) many body configuration are created at the Fermi level in the correlation gap by doping  (Bianconi \& Congiu-Castellano  \& De-Santis \& Rudolf, \textit{et al.}, 1987) \cite{BianconiRudolf1987},  
(Bianconi \& Clozza \& Congiu-Castellano \& Della-Longa \textit{et al.}, 1987) \cite{BianconiClozza1987}, (Bianconi \& Clozza, \& Congiu-Castellano, \& Della Longa, \&  De Santis, \&  Di Cicco, \textit{et al.}, 1987)  \cite{BianconiClozzaDiCicco1987} and these results were presented at the Taniguchi symposium, Kashikojima, Japan, october 19-23, 1987.(Kanamori \& Kotani, 1988)   \cite{KanamoriKotani1988}. In 1988  these results have been confirmed by similar findings in other cuprates (Bianconi \& Budnick \& Chamberland \textit{et al.},  1988)   \cite{BianconiBudnickChamberland1988} and in formally trivalent Cu compounds (Bianconi \& Budnick \& Demazeau \textit{et al.}, 1988)  \cite{BianconiBudnickDemazeau1988}.
While other authors presented different interpretations of the XANES data in 1987 (Grioni \textit{et al.}, 1987)  \cite{Grioni1987}, (Nuecker \textit{et al.}, 1987)  \cite{Nuecker1987}, the oxygen K-edge  absorption  (Nuecker  \textit{et al.}, 1988) \cite{Nuecker1988} provided a compelling evidence for the formation of  Cu$3d^9$L  many body states induced by doping. 
 
 In 1988 using polarized XANES spectroscopy of single crystals it was shown that there are two coexisting different many body configurations:  Cu$3d^9$L ($b_1$), where the holes on oxygen orbital have fully planar symmetry L($b_1$)  and and Cu$3d^9$L($a_1$) where the oxygen holes have partially out-of-plane symmetry  L($a_1$)  (Bianconi \& Desantis \& Flank \textit{et al.}, 1988)  \cite{BianconiDesantisFlank1988}, (Bianconi \& De Santis \& Di Cicco \textit{et al.}, 1988) \cite{BianconiDeSantisDiCicco1988}, 
 (Bianconi \&  De Santis \& Di Cicco \textit{et al.}, 1989) \cite{BianconiDeSantisDiCicco1989}. 
  
The presence of two states with different orbital symmetry is relevant for superconductivity (Flank \textit{et al.}, 1990)   \cite{Flank1990}.
 The presence of a second electronic component Cu$3d^9$L($a_1$) was shown to be associated with the rhombic distortion on the $CuO_4$ square plane (Seino  \textit{et al.},1990) \cite{Seino1990} related to charge density wave onset and polaron formation (Bianconi  \&   Missori \&  Oyanagi, \textit{et al.}, 1995)  \cite{BianconiMissoriOyanagi1995}. 
 
 Full multiple scattering analysis of polarized K-edge  (Li  \textit{et al.}, 1991) \cite{Li},  (Bianconi \&  Li \& Campanella, \textit{et al.}, 1991)  \cite{BianconiLiCampanella1991},   
 and  Cu $L_3$-edge (Pompa \textit{et al.}, 1991)  \cite{Pompa 1991},
 (Bianconi \&  Della Longa \& Li \textit{et al.}, 1991) \cite{BianconiDellaLongaLi1991} 
 XANES of $La_2CuO_4$ and $Bi_2CaSr_2Cu_2O_{8+y}$ have clearly shown that many body final states configurations give a splitting of the strongest peak in the Cu K-edge XANES.
  The presence of two set of states at the Fermi level with different orbital character has been confirmed by many experiments at O 1s and Cu 2p edges in  $La_{2-x}Sr_xCuO_{4+y}$, $La_{2-x}Sr_xNiO_{4+y}$, $(Y,R)Ba_2Cu_3O_7$  
 (Pellegrin \textit{et al.}, 1993)  \cite{Pellegrin1993},
  $Bi_2CaSr_2Cu_2O_{8+y}$ 
 (Pellegrin \textit{et al.}, 1995)  \cite{Pellegrin1995} 
 and $HgBa_2Ca_{n-1}Cu_nO_{2n+2+y}$ 
 (Pellegrin \textit{et al.}, 1996)   \cite{Pellegrin1996},  $Pr_xY_{1-x}Ba_2Cu_3O_{7-y}$ 
 (Merz, \textit{et al.}, 1997) \cite{Merz1997}.  Today the  two electronic many body components seen by XANES provide the experimental base for the recent development of the theory of multigap superconductivity both in cuprates  
 (Valletta \textit{et al.}, 1997,Bianconi textit{et al.}, 1997) \cite{Valletta1997,bianconi97}
 (Perali \textit{et al.}, 2012) \cite{Perali2012} and other high temperature superconductors including pressurized sulfur hydrides \cite{h3s1,h3s2,h3s3.h3s4}.

\end{document}